\documentclass[11pt, a4paper]{article}
\usepackage[text={8in,11in},margin=1.0in,includefoot]{geometry}
\usepackage{secdot}
\sectiondot{subsection}
\usepackage{doc}
\usepackage{booktabs,caption}
\usepackage{array}
\newcolumntype{P}[1]{>{\centering\arraybackslash}p{#1}}
\newcolumntype{M}[1]{>{\centering\arraybackslash}m{#1}}

\usepackage{url}
\usepackage{graphicx}
\usepackage{epstopdf}
\usepackage{tabularx}
\usepackage{amsmath}
\usepackage{pdflscape}
\usepackage{float}
\usepackage{placeins}
\usepackage{caption}
\usepackage{booktabs}
\usepackage{cite}
\usepackage{multirow}
\usepackage{longtable}
\usepackage[export]{adjustbox}
\usepackage[english]{babel}
\usepackage[autostyle]{csquotes}

\restylefloat{table}
\numberwithin{equation}{section}
\title{$J^{P}=\frac{1}{2}^{+}, \frac{3}{2}^{+}$ masses in statistical model }
\author{$^1$Amanpreet Kaur,$^2$Alka Upadhyay
\\\small{\it School of Physics and Material Science},\\\small{\it Thapar University,
Patiala, Punjab-147004}\\\small{E-mail:
$^1$amanpreet.kaur9074@yahoo.com, $^2$alka.iisc@gmail.com}}
\begin{document}
\date{}

\maketitle
\begin{abstract} The mass formulae for the baryon octet and decuplet are calculated. These formulae are function
of constituent quark masses and spin spin interaction terms for the
quarks inside the baryons. The coefficients in the mass formulae is
estimated by the statistical model for $J^{P}=1/2^{+}, 3/2^{+}$,
incorporating the contributions from \enquote{sea} containing
$u\overline{u}, d\overline{d}, s\overline{s}$ pairs and gluons . The
measured masses are presented and found to be matching good with
some of the experimental and theoretical data.

\vspace{0.2cm} \noindent {\bf  }

\end{abstract}
\section{Introduction}
Many calculations have been performed in the last few years with the
aim of understanding the baryon mass spectrum. In order to
understand the hadron spectroscopy the low energy properties should
be essentially understood. A simple yet unique concept of particle
physics, i.e. quark model suggest that all hadrons are made of of
three quarks or a quark and an antiquark, bounded by all
interactions which arise from renormalizable gauge couplings. In
this paper, we will try to understand the baryonic mass spectrum
from this standpoint. Many of the hadron mass splitting observed
till date are produced by the splitting among the quark masses. Much
research has been devoted in describing hadron masses which also
incudes masses of hadrons with heavy quarks, often with considerable
success \cite{1,2,3,4,5,6,7,8,9}. Morpurgo \cite{10}, by the use of
field theory and the non relativistic quark model, gave a general
parameterizations of the magnetic moments and masses of baryon octet
and decuplet. His general parametrization method (GPM), derived from
the QCD Lagrangian, expresses the mass operators in terms of
flavor-dependent terms proportional to powers of the strange-quark
projection operator $P^{\lambda}$ and non relativistic appearing
products of Pauli spin operators $\sigma$.

Several authors have studied the masses of heavy hadrons in the
non-relativistic quark model with the inclusion of spin and flavor
dependant hyperfine interaction between two quarks and between a
quark and an antiquark and many other different techniques
\cite{12}. Study of baryon mass spectrum has been a subject of
increasing interest due to related experiments at Fermilab, CERN
etc. \cite{13}. Also several models including non relativistic quark
model (NRQM) \cite{2,4}, chiral perturbation theory (ChPT)
\cite{14}, hyper central model \cite{15} have evaluated the data of
light and heavy baryon mass spectrum having nice agreement with
experimental information available. C. P. Singh \cite{2} et. al.,
calculated the masses of charmed and b-quark hadrons in the non
relativistic additive quark model with the inclusion of spin and
flavor dependant hyperfine interaction between two quarks and a
quark and an antiquark, which was in agreement with the available
data.

Masses of heavy and light baryons are computed using the equation
Mass = $M_{quark}$ + Spin-Spin Correction, where $M_{quark}$ is the
mass obtained using scalar quarks and the spin-spin correction term.
In order to understand mass spectrum, spin-spin interaction term has
a pivotal role in the mass formulae. This term is short range, in
the sense that interaction energy associated with spin-spin coupling
of quarks decreases with distance as $\alpha_{S}^{-N}$. In general,
it is only the short range forces between the quarks which are spin
dependant. On the other hand, the lattice is good for the long
distance physics, the lattice cutoff can spoil the short-distance
physics. In this note, we consider the baryon masses in the
statistical picture where the short range forces are focussed. It is
worthy to mention that we are considering S-wave quark-antiquark
systems here. The parameters required for these calculations are
completely determined by previously studied properties, using the
wave functions, so the results are entirely predictive. The
spin-spin interaction contribution is also responsible for
$\Sigma-\Lambda$ mass splitting in case of baryons \cite{11}. The
reason for this splitting is that, in case of $\Sigma^{0}$, the
quarks are in symmetric state which means that quark pair (u,d) must
also be symmetric in spin state, such that
$\overrightarrow{S}_{u}.\overrightarrow{S}_{d}=\frac{1}{4}$. For
$\Lambda$, the quark pair (u,d) has isospin zero which means total
spin is zero (antisymmetric state), such that
$\overrightarrow{S}_{u}.\overrightarrow{S}_{d}=\frac{-3}{4}$. The
\enquote{hyperfine} splittings for $\Sigma-\Lambda$ are related by:
\begin{gather*}
\Sigma-\Lambda=\frac{2}{3}(1-\frac{m_{u}}{m_{d}})(\Delta-N)
\end{gather*}
The scheme of the the present manuscript is as follows: Section II
begins with a brief review of construction of decuplet wave function
with sea. Section III presents the explanation of model used i.e.
statistical model and principle of detailed balance. Mass formulae
are defined in section IV followed by calculation of the masses in
section V. Discussion and conclusion is presented in Section VI.

\section{Preliminaries} The structure of hadron constitutes two parts i.e valence
part (qqq) and other is sea part which consist of quark-antiquark
pairs muticonnected through gluons \cite{16,17,18}. A $q^{3}$ state
in the baryon are in the 1, 8 and 10 color states which means that
sea should also be in corresponding states to form a color singlet
baryon. The valence part of the hadronic wave function can be
written as:
\begin{equation}
    \label{eq:wavefunction}
  \Psi=\Phi(|\phi\rangle|\chi\rangle|\psi\rangle)(|\xi\rangle)
     \end{equation}where $|\phi\rangle,|\chi\rangle,|\psi\rangle$ and $|\xi\rangle$
denote flavor, spin, color and space $q^3$ wave functions and their
contribution yields antisymmetric total wave function. Here, spatial
part $(|\xi\rangle)$ is symmetric under the exchange of any two
quarks for the hadrons and therefore the flavor-spin-color part
$\Phi(|\phi\rangle|\chi\rangle|\psi\rangle)$ should be antisymmetric
in nature such that when combined with $(|\xi\rangle)$ gives
antisymmetric total wave function. To show an active participation
of sea, a relevant wave function is written with valence and quark
gluon Fock states. Sea considered here is in S-wave state with spin
(0,1,2) and color (1,8,$\overline{10}$) and is assumed to be
flavorless. Let $H_{0,1,2}$ and $G_{1,8,\overline{10}}$ denote spin
and color sea wave functions, which satisfy $\langle
H_{i}|H_{j}\rangle=\delta_{ij}$ , $\langle
G_{k}|G_{l}\rangle=\delta_{kl}$. In this approach we have assumed a
sea to be consisting of three gluons. All possible combinations of
valence q$^3$ and sea wave functions which can yield spin 1/2 (3/2),
flavor octet (decuplet) and color singlet state thereby maintaining
the anti symmetrization of the total baryonic wave function are
\cite{19,20}:

\begin{equation}
\begin{split}
\label{eq:wavefunction1} \text{Octet}= \Phi_{1}^{(1/2)}H_{0}G_{1} ,
\Phi_{8}^{(1/2)}H_{0}G_{8},
 \Phi_{10}^{(1/2)}H_{0}G_{\overline{10}},
\Phi_{1}^{(1/2)}H_{1}G_{1} ,\Phi_{8}^{(1/2)}H_{1}G_{8},\\
\Phi_{10}^{(1/2)}H_{1}G_{\overline{10}}, \Phi_{8}^{(3/2)}H_{1}G_{8},
\Phi_{8}^{(3/2)}H_{2}G_{8}
\end{split}
\end{equation}

\begin{equation}
\begin{split}
\label{eq:wavefunction1}
 \text{Decuplet}=\Phi_{1}^{(3/2)}H_{0}G_{1} ,\Phi_{1}^{(3/2)}H_{1}G_{1} ,
\Phi_{8}^{(1/2)}H_{1}G_{8}  ,\Phi_{1}^{(3/2)}H_{2}G_{1},
\Phi_{8}^{(1/2)}H_{2}G_{8}
\end{split}
\end{equation}
The total flavor-spin-color wave function of a spin up baryon octet
(decuplet) consisting of three valence quarks and a sea component
can be written as:

\begin{equation}
\begin{split}
\label{eq:wavefunction2} \text{Octet}=
|\Phi_{1/2}^{(\uparrow)}\rangle=\frac{1}{N}[\Phi_{1}^{(1/2\uparrow)}H_{0}G_{1}+
a_{8}(\Phi_{8}^{(1/2)}\otimes H_{0})^{\uparrow}G_{8}+a_{10}\Phi_{10}^{(1/2\uparrow)}H_{0}G_{\overline{10}}\\
b_{1}(\Phi_{8}^{(1/2)}\otimes
H_{1})^{\uparrow}G_{1}+b_{8}(\Phi_{8}^{(1/2)}\otimes
H_{1})^{\uparrow}G_{8}+b_{10}(\Phi_{10}^{(1/2)}\otimes
H_{1})^{\uparrow}G_{\overline{10}}\\+ c_{8}(\Phi_{8}^{(3/2)}\otimes
H_{1})^{\uparrow}G_{8}+ d_{8}(\Phi_{8}^{(3/2)}\otimes
H_{2})^{\uparrow}G_{8}]
\end{split}
\end{equation}
where
\begin{equation}
\label{eq:wavefunction3}N^{2}=1+a_{8}^{2}+a_{10}^{2}+b_{1}^{2}+b_{8}^{2}+b_{10}^{2}+c_{8}^{2}+d_{8}^{2}
\end{equation}

\begin{equation}
\begin{split}
\label{eq:wavefunction2} \text{Decuplet}=
|\Phi_{3/2}^{(\uparrow)}\rangle=\frac{1}{N}[a_{0}\Phi_{1}^{(3/2\uparrow)}H_{0}G_{1}+
b_{1}(\Phi_{1}^{(3/2)}\otimes H_{1})^{\uparrow}G_{1}+\\
b_{8}(\Phi_{8}^{(1/2)}\otimes H_{1})^{\uparrow}G_{8}+
d_{1}(\Phi_{1}^{(3/2)}\otimes H_{2})^{\uparrow}G_{1}+\\
d_{8}(\Phi_{8}^{(1/2)}\otimes H_{2})^{\uparrow}G_{8}]
\end{split}
\end{equation}
where
\begin{equation}
\label{eq:wavefunction3}N^{2}=a_{0}^{2}+b_{1}^{2}+b_{8}^{2}+d_{1}^{2}+d_{8}^{2}
\end{equation}
Here, N is the normalization constant. The first three terms in the
eq. (2.4) are obtained by combining $q^{3}$ wave function with spin
0 (scalar sea) and next three terms are obtained by coupling $q^{3}$
with spin 1 (vector sea). The final two terms are the result of
coupling with spin 2 (tensor sea). Similarly, the first term in eq.
(2.6) is obtained by combining the $q^{3}$ wave function with spin 0
(scalar sea) and the next two terms are obtained by coupling $q^{3}$
with spin 1 (vector sea) and the final two terms are the result of
the coupling with spin 2 (tensor sea). The details of all the terms
of wave function in equation (2.4) and (2.6) can be found in
references \cite{19,20}. All the coefficients in the above wave
functions are determined statistically from the flavor, spin and
color probabilities for the study of low energy properties of
hadrons.

\section{Principle of detailed balance and Statistical Model}
The main idea of the detailed balance model, proposed by Zhang et.
al. \cite{21,22}, is that it assumes the proton as a bag of
quark-gluon gas in dynamical balance, where partons keep combining
and splitting through processes such as $g\Leftrightarrow
q\overline{q}$,$g\Leftrightarrow gg$,$q\Leftrightarrow qg$. This
model was proposed to study the $\overline{d}-\overline{u}$
asymmetry in nucleon and it was found that the detailed balance
model gives $\overline{d}-\overline{u}$=0.124 \cite{23}, which was
in agreement with the E866/NuSea result of 0.118 \cite{24}. This
good agreement indicates that the principle of detailed balance
plays an important role in the structure of proton. The method was
also extended to pions \cite{25} and the nucleon spin structure
\cite{26}. Later, the strange content of the proton was also
calculated, using the balance model under the equal probability
assumption.

The model begins with the valance quark structure of the proton
without any parameters. In this picture, while $d\overline{d}$ and
$u\overline{u}$ sea quark-antiquark pairs are produced by gluon
splitting with equal probability, the reverse process, i.e., the
annihilation of antiquarks with their quark partners into gluons, is
not flavor symmetric due to excess of u quarks over d quarks in the
proton. In general, detailed balance principle demands that the
exchange between any two states should balance each other, which can
be expressed as:
\begin{equation}
n_{A\rightarrow B}=n_{B\rightarrow A}
\end{equation} where A and B are the states.
Hadron is treated to be consisting of complete set of quark gluon
Fock states and can be expressed in expanded form as:
\begin{equation}
\label{eq:wavefunction23}
|Baryon\rangle=\sum_{i,j,l,k}C_{i,j,l,k}|(q^{3}),(i,j,l),(k)\rangle
\end{equation}
where ${q^{3}}$ represents the valence quarks of the baryon , i is
the number of $u\overline{u}$ pairs, j is the number of
$d\overline{d}$ pairs, {l} is the number of $s\overline{s}$ pairs
and k is the number of gluons in sea. The probability to find a
quark-gluon Fock states in the baryon system is:
\begin{equation}
\label{eq:wavefunction24} \rho_{i,j,l,k}=|C_{i,j,l,k}|^{2},
\end{equation} and
$\rho_{i,j,l,k}$ satisfies the normalization condition,
\begin{equation}
\label{eq:wavefunction25} \sum_{i,j,l,k}\rho_{i,j,l,k}=1
\end{equation}
The transition probabilities in flavor space for various Fock states
have already been determined for nucleon and hyperons containing
strange sea and can be found in references \cite{22,27}. This model,
till date, has been able to explain flavor asymmetry, magnetic
moments, spin distribution, semileptonic decays of nucleon for octet
and decuplet particles \cite{27,28,20,20a}. We will, here, use this
model to calculate the mass splittings of octet and decuplet
particles. Also, it is worthy to mention that the quarks and gluons
in the Fock states are the "intrinsic" partons of hadrons as they
are non-perturbatively multiconnected to valence quarks. The
\enquote{intrinsic} sea quarks and gluons survive over a long
lifetime within hadronic bound states whereas the
\enquote{extrinsic} sea quarks and gluons only exist for a short
time. The "extrinsic" partons in the Fock satates are generated from
QCD hard bremsstrahlung or gluon splitting as part of the lepton
scattering interaction.

\medskip {\bf The Statistical model} \cite{26} is used in our formalism to calculate the masses of octet and decuplet
members by assuming hadrons as an ensemble of three valence quarks
and sea containing various quark-gluon Fock states. The main feature
of the statistical model is that it does not requires any additional
input parameters which proves its physical simplicity, that have
made an amazing success in describing parton distribution functions
for nucleons. A more general description of this model was provided
by J.P Singh et. al., where in addition to flavor, each Fock state
has some definite spin and color quantum number with a specific
symmetry property \cite{26}. Here, it is worthy to mention that all
the above listed properties were directly linked to probabilities of
each Fock state in definite spin, color, flavor space quantum
numbers. The different possible states in spin, flavor and color,
for two gluons can be written as:
\begin{gather*}
\textbf{Spin}: gg:1\otimes1=0_{s}\oplus1_{a}\oplus2_{s}\\
\textbf{Color}: gg:
8\otimes8=1_{s}\oplus8_{s}\oplus8_{a}\oplus10_{a}\oplus\overline{10}_{a}\oplus27_{s}\\
\textbf{Flavor}:
3\otimes3\otimes3=1_{a}\oplus8_{ms}\oplus8_{ma}\oplus10_{s}
\end{gather*}
The decuplet is symmetric in flavor, singlet antisymmetric and the
two octets have mixed symmetry. Subscripts s and a denotes symmetry
and antisymmetry on combining the states. Total antisymmetry of the
baryon should be kept in mind while combining the valence and sea
part. In this model, all $n_{\mu,\nu}'s$ are calculated from
multiplicities of each Fock state in spin and color space. These
multiplicities are expressed in the form of $\rho_{p,q}$ where
relative probability for core part should have spin p and sea to
have spin q such that the resultant should come out as 1/2 (3/2).
Similar probabilities could be calculated for color space which
yields color singlet state. Calculation of these probabilities helps
to find common factor \enquote{c} for every combination of valence
and sea which is multiplied with multiplicity factor (n) for each
Fock state. The common parameter \enquote{c} can be calculated from
the various Fock states derived from the principle of detailed
balance. Each unknown parameter in the equation of wave function
will have a particular value of $\sum n_{\mu\nu}c_{sea}$ depending
on the Fock state \cite{20}:
\begin{eqnarray}
a_{0}^{2}=(n_{01}c_{sea})_{|gg\rangle}+(n_{01}c_{sea})_{|u\overline{u}g\rangle}+(n_{01}c_{sea})_{|d\overline{d}g\rangle}+(n_{01}c_{sea})_{|s\overline{s}g\rangle}+...\\b_{1}^{2}=(n_{11}c_{sea})_{|gg\rangle}+(n_{11}c_{sea})_{|u\overline{u}g\rangle}+(n_{11}c_{sea})_{|d\overline{d}g\rangle}+(n_{11}c_{sea})_{|s\overline{s}g\rangle}+...\\d_{1}^{2}=(n_{21}c_{sea})_{|gg\rangle}+(n_{21}c_{sea})_{|u\overline{u}g\rangle}+(n_{21}c_{sea})_{|d\overline{d}g\rangle}+(n_{21}c_{sea})_{|s\overline{s}g\rangle}+...\\............................................................................................................
\end{eqnarray}
Combinations for other unknown parameters can be written in a
similar way. A detailed information for calculation of all the
parameters using statistical model is being provided in references
\cite{19,20}. The calculations for probabilities are done in two
ways i.e C model and D model. Model D assumes that a sea containing
a large number of gluons has relatively smaller probabilities and
hence their higher multiplicities have been suppressed over the rest
of valence particles with limited quarks. Sea with larger color
multiplicity has less probability of survival due to larger
possibility of interaction. Model D is assumed to be a special case
of Model C. The parametrization of model D, can be achieved by
assuming that probability of a state is inversely proportional to
multiplicity of state in both spin and color states. Therefore, the
new probabilities are additional to previous found probabilities
factors.

\section{Mass Formulae}
The hadronic mass spectrum is an essential ingredient in theoretical
study of the physics involving strong interactions. Mass spectrum of
mesons and baryons are studied in various models and aspects. The
hadron mass formula is discussed in the quark-counting aspect which
shows that the \enquote{free} quark picture gives the GellMann-Okubo
formula \cite{28a}. In quark models, the baryon octet and decuplet
are bound state of three-quark states ($q^{3}$)and there are various
different model calculations for thier masses. Empirical mass
formulae \cite{28b} for the baryon octet and decuplet are the
functions of one integer variable assigned to each member of the
baryon sets and charge state of the baryon. These formulae are
independent of any specific model. Further, the gross features of
simple quark model has helped in unraveling the detailed properties
of mass spectra of baryons and mesons. For example, how we determine
the different masses of $\Lambda^{(\frac{1}{2})}$(1115.683),
$\Sigma^{0(\frac{1}{2})}$ (1192.642) and $\Sigma^{*0(\frac{3}{2})}$
(1382.8), despite having same quark content. The difference in spin
spin interactions among quarks is the answer to this query, as
explained in section 1. We assume, that mass of hadron arises from
the constituent quark masses plus the spin-spin interaction energies
of quarks for a meson and a baryon, can be written as \cite{29}:
\begin{equation}
M_{meson}=a_{m}+m_{i}+m_{j}+bm_{0}^{2}\frac{s_{i}.s_{j}}{m_{i}m_{j}}
\end{equation}

\begin{equation}
M_{baryon}=a_{b}+\sum \limits_{\substack{i}}
m_{i}+\frac{bm_{0}^{2}}{3}\sum \limits_{\substack{i<j}}
\frac{s_{i}.s_{j}}{m_{i}m_{j}}
\end{equation}
where $a_{m}$, $a_{b}$, b are parameters with the dimension of mass
and $m_{i}$, $m_{j}$ are the masses of respective quarks
(antiquarks), $m_{0}$ is a scale factor which we shall take to be
the mass of the lightest quark, i.e. $m_{0}= m_{u},m_{d}$. The spin
dependent term includes a contribution of each pair that is
proportional to the expectation value of $s_{i}.s_{j}$ and inversely
proportional to the product of the constituent quark masses
$m_{i}m_{j}$. The spin-spin interaction used in the mass operator
may be interpreted as the interaction between color magnetic moments
proportional to g/2m in analogy with Dirac magnetic moment, g is the
strong-coupling constant of gluons with quarks. With the help of the
wave function of baryon octet and decuplet, described in previous
section, eqns. (3.1) and (3.2) give the masses of the hadrons in
terms of the parameters. To evaluate equation (4.2) for each baryon,
we need to find the expectation values for spin operators for each
quark pair within the respective baryon. The spin interaction term
we need to find for these baryons, which are made of up of, say, u,
d, s is,
\begin{equation}
\frac{\overrightarrow{S}_{u}.\overrightarrow{S}_{d}}{m_{u}^{2}}+\frac{\overrightarrow{S}_{u}.\overrightarrow{S}_{s}+\overrightarrow{S}_{d}.\overrightarrow{S}_{s}}{m_{u}m_{s}}
\end{equation}Here, $m_{u}=m_{d}$. The eigen values for $s_{u}.s_{d}$ are 1/4 and -3/4 for spin triplet
(I=1) and singlet (I=0) states respectively, results in evaluation
of other terms by: Consider
$\overrightarrow{J}=\overrightarrow{S}_{u}+\overrightarrow{S}_{d}+\overrightarrow{S}_{s}$
\begin{equation}
\text{Then},
\overrightarrow{S}_{u}.\overrightarrow{S}_{s}+\overrightarrow{S}_{d}.\overrightarrow{S}_{s}=\frac{1}{2}[J^{2}-(S_{u}^{2}+S_{d}^{2}+S_{s}^{2})]-\overrightarrow{S}_{u}.\overrightarrow{S}_{u}
\end{equation}

\begin{equation}
\begin{split}
\text{For octet},
\overrightarrow{S}_{u}.\overrightarrow{S}_{s}+\overrightarrow{S}_{d}.\overrightarrow{S}_{s}=
-1 (\text{symmetric})\\
\overrightarrow{S}_{u}.\overrightarrow{S}_{s}+\overrightarrow{S}_{d}.\overrightarrow{S}_{s}=
0 (\text{antisymmetric})
\end{split}
\end{equation}

\begin{equation}
\begin{split}
\text{For decuplet},
\overrightarrow{S}_{u}.\overrightarrow{S}_{s}+\overrightarrow{S}_{d}.\overrightarrow{S}_{s}=
\frac{1}{2} (\text{symmetric})\\
\overrightarrow{S}_{u}.\overrightarrow{S}_{s}+\overrightarrow{S}_{d}.\overrightarrow{S}_{s}=
\frac{3}{2}(\text{antisymmetric})
\end{split}
\end{equation}
These values are applicable for all the baryon octet and decuplet
particles relative to their quark content. The mass operator given
in equation (4.2) is applied to the terms of the wave function in
equation (2.4) and (2.6). The eigen values of spin operator given in
equation (4.5) and (4.6) is then used to obtain the relations in
terms of parameters with the dimension of mass. Mass relations thus
obtained are displayed in table 1 and 2 for $J^{P}=\frac{1}{2}^{+}$,
$\frac{3}{2}^{+}$ states.
\setlength{\tabcolsep}{0.9 em} %
{\renewcommand{\arraystretch}{1.1}%
\begin{table}[H]
\caption{Derived mass relation for $J^{P}=\frac{1}{2}^{+}$
particles. Here, $m_{i}$, $m_{j}$ are the constituent quark masses
of the respective quarks in the baryon, $a_{b}$, b are the
parameters with the dimension of mass and $m_{0}$ is a scale factor
equal to mass of the lightest quark, i.e $m_{0}=m_{u}, m_{d}$.}
\begin{center}
\begin{tabular}{|l|p{11cm}|}
\hline Term & After applying operator on each term\\\hline
$a_{0}\Phi_{1}^{(1/2)}H_{0}G_{1}$& $a_{0}(a_{b}+\sum
\limits_{\substack{i}}
m_{i}+\frac{bm_{0}^{2}}{3}(\frac{1}{\sqrt{2}})[\frac{1}{4m_{i}m_{j}}-\frac{1}{m_{i}m_{j}}-\frac{3}{4m_{i}m_{j}}])$
\\\hline

$a_{8}\Phi_{8}^{(1/2)}H_{0}G_{8}$& $a_{8}(a_{b}+\sum
\limits_{\substack{i}}
m_{i}+(-\frac{bm_{0}^{2}}{3}[\frac{1}{4m_{i}m_{j}}-\frac{1}{m_{i}m_{j}}]))$
\\\hline

$a_{10}\Phi_{10}^{(1/2)}H_{0}G_{\overline{10}}$& $a_{10}(a_{b}+\sum
\limits_{\substack{i}}
m_{i}+\frac{bm_{0}^{2}}{3}(\frac{1}{\sqrt{2}})[-\frac{3}{4m_{i}m_{j}}-\frac{1}{4m_{i}m_{j}}+\frac{1}{m_{i}m_{j}}])$
\\\hline

$b_{1}(\Phi_{1}^{(1/2)}\otimes H_{1})G_{1}$& $b_{1}(a_{b}+\sum
\limits_{\substack{i}}
m_{i}+\frac{bm_{0}^{2}}{3}(\frac{1}{\sqrt{3}}-\frac{1}{\sqrt{6}})[-\frac{3}{4m_{i}m_{j}}+\frac{1}{4m_{i}m_{j}}-\frac{1}{m_{i}m_{j}}])$
\\\hline

$b_{8}(\Phi_{8}^{(1/2)}\otimes H_{1})G_{8}$& $b_{8}(a_{b}+\sum
\limits_{\substack{i}}
m_{i}+\frac{bm_{0}^{2}}{3}(\frac{1}{\sqrt{6}}-\frac{1}{\sqrt{12}})[\frac{1}{4m_{i}m_{j}}-\frac{1}{m_{i}m_{j}}+\frac{3}{4m_{i}m_{j}}]+\newline
(\frac{1}{\sqrt{12}}-\frac{1}{\sqrt{6}})[-\frac{3}{4m_{i}m_{j}}+\frac{1}{4m_{i}m_{j}}-\frac{1}{m_{i}m_{j}}])$
\\\hline

$b_{10}(\Phi_{10}^{(1/2)}\otimes H_{1})G_{\overline{10}}$&
$b_{10}(a_{b}+\sum \limits_{\substack{i}}
m_{i}+\frac{bm_{0}^{2}}{3}(\frac{1}{\sqrt{3}}-\frac{1}{\sqrt{6}})[\frac{1}{4m_{i}m_{j}}-\frac{1}{m_{i}m_{j}}+\frac{3}{4m_{i}m_{j}}])$
\\\hline

$c_{8}(\Phi_{8}^{(3/2)}\otimes H_{1})G_{8}$& $c_{8}(a_{b}+\sum
\limits_{\substack{i}}
m_{i}+\frac{bm_{0}^{2}}{3}(\frac{1}{2}-\frac{1}{\sqrt{6}}+\frac{1}{\sqrt{12}})[\frac{3}{4m_{i}m_{j}}])$
\\\hline

$d_{8}(\Phi_{8}^{(3/2)}\otimes H_{2})G_{8}$& $d_{8}(a_{b}+\sum
\limits_{\substack{i}}
m_{i}+\frac{bm_{0}^{2}}{3}(\frac{1}{\sqrt{5}}-\sqrt{\frac{3}{20}}+\frac{1}{\sqrt{10}}-\frac{1}{\sqrt{20}})[\frac{1}{4m_{i}m_{j}}-\frac{1}{m_{i}m_{j}}])$
\\\hline

\end{tabular}
\end{center}
\end{table}

\setlength{\tabcolsep}{1.1em} %
{\renewcommand{\arraystretch}{1.5}%
\begin{table}[H]
\caption{Derived mass relation for $J^{P}=\frac{3}{2}^{+}$
particles. Here, $m_{i}$, $m_{j}$ are the constituent quark masses
of the respective quarks in the baryon, $a_{b}$, b are the
parameters with the dimension of mass and $m_{0}$ is a scale factor
equal to mass of the lightest quark, i.e $m_{0}=m_{u}, m_{d}$. }
\begin{center}
\begin{tabular}{|l|p{10cm}|}
\hline Term & After applying operator on each term \\\hline
$a_{0}\Phi_{1}^{(3/2)}H_{0}G_{1}$& $a_{0}(a_{b}+\sum
\limits_{\substack{i}}
m_{i}+\frac{bm_{0}^{2}}{3}[\frac{1}{4m_{i}^{2}}+\frac{1}{2m_{i}m_{j}}])$
\\\hline

$b_{1}\Phi_{1}^{(3/2)}H_{1}G_{1}$&$b_{1}( a_{b}+\sum
\limits_{\substack{i}}
m_{i}+\frac{bm_{0}^{2}}{3}[\sqrt{\frac{3}{5}}[\frac{-3}{4m_{i}^{2}}+\frac{3}{2m_{i}m_{j}}]-\sqrt{\frac{2}{5}}[\frac{-3}{4m_{i}^{2}}+\frac{3}{2m_{i}m_{j}}]])$\\\hline

$b_{8}\Phi_{8}^{(1/2)}H_{1}G_{8}$&$b_{8}(a_{b}+\sum
\limits_{\substack{i}}
m_{i}+\frac{bm_{0}^{2}}{3}[\sqrt{\frac{3}{5}}[\frac{-3}{4m_{i}^{2}}+\frac{3}{2m_{i}m_{j}}]-\sqrt{\frac{3}{5}}[\frac{-3}{4m_{i}^{2}}+\frac{3}{2m_{i}m_{j}}]])$\\\hline

$d_{1}\Phi_{1}^{(3/2)}H_{2}G_{1}$&$d_{1}(a_{b}+\sum
\limits_{\substack{i}}
m_{i}+\frac{bm_{0}^{2}}{3}[\sqrt{\frac{1}{5}}[\frac{1}{4m_{i}^{2}}+\frac{1}{2m_{i}m_{j}}]-\sqrt{\frac{2}{5}}[\frac{-3}{4m_{i}^{2}}+\frac{3}{2m_{i}m_{j}}]+
\sqrt{\frac{2}{5}}[\frac{-3}{4m_{i}^{2}}+\frac{3}{2m_{i}m_{j}}]])$\\\hline

$d_{8}\Phi_{8}^{(1/2)}H_{2}G_{8}$&$d_{8}(a_{b}+\sum
\limits_{\substack{i}}
m_{i}+\frac{bm_{0}^{2}}{3}[\sqrt{\frac{1}{10}}[\frac{1}{m_{i}^{2}}-\frac{1}{m_{i}m_{j}}]-\sqrt{\frac{2}{10}}[\frac{1}{m_{i}^{2}}-\frac{1}{m_{i}m_{j}}]+
\sqrt{\frac{2}{10}}[\frac{1}{m_{i}^{2}}-\frac{1}{m_{i}m_{j}}]])$\\\hline
\end{tabular}
\end{center}
\end{table}

The calculation of mass for proton is shown below:
\begin{equation}
\begin{split}
Mass_{(proton)} =0.43(a_{b}+m_{u}+m_{u}+m_{d}+
\frac{bm_{0}^{2}}{3}(\frac{1}{\sqrt{2}})[\frac{1}{4m_{u}m_{u}}-\frac{1}{m_{u}m_{d}}-\frac{3}{4m_{u}m_{u}}])+\\
0.022(a_{b}+m_{u}+m_{u}+m_{d}+(-\frac{bm_{0}^{2}}{3}[\frac{1}{4m_{u}m_{u}}-\frac{1}{m_{u}m_{d}}]))+\\
0.003(a_{b}+m_{u}+m_{u}+m_{d}+
\frac{bm_{0}^{2}}{3}(\frac{1}{\sqrt{2}})[-\frac{3}{4m_{u}m_{u}}-\frac{1}{4m_{u}m_{u}}+\frac{1}{m_{u}m_{d}}])+\\0.142
(a_{b}+m_{u}+m_{u}+m_{d}+\frac{bm_{0}^{2}}{3}(\frac{1}{\sqrt{3}}-\frac{1}{\sqrt{6}})[-\frac{3}{4m_{u}m_{u}}+
\frac{1}{4m_{u}m_{u}}-\frac{1}{m_{u}m_{d}}])+\\0.014
(a_{b}+m_{u}+m_{u}+m_{d}+\frac{bm_{0}^{2}}{3}(\frac{1}{\sqrt{6}}-\frac{1}{\sqrt{12}})[\frac{1}{4m_{u}m_{u}}-\frac{1}{m_{u}m_{d}}+\frac{3}{4m_{u}m_{u}}]+\\
(\frac{1}{\sqrt{12}}-\frac{1}{\sqrt{6}})[-\frac{3}{4m_{u}m_{u}}+\frac{1}{4m_{u}m_{u}}-\frac{1}{m_{u}m_{d}}])+\\0.0023
(a_{b}+m_{u}+m_{u}+m_{d}+\frac{bm_{0}^{2}}{3}(\frac{1}{\sqrt{3}}-\frac{1}{\sqrt{6}})[\frac{1}{4m_{u}m_{u}}-\frac{1}{m_{u}m_{d}}+\frac{3}{4m_{u}m_{u}}])+\\0.0035
(a_{b}+m_{u}+m_{u}+m_{d}+\frac{bm_{0}^{2}}{3}(\frac{1}{2}-\frac{1}{\sqrt{6}}+\frac{1}{\sqrt{12}})[\frac{3}{4m_{u}m_{u}}])+\\0.0014
(a_{b}+m_{u}+m_{u}+m_{d}+\frac{bm_{0}^{2}}{3}(\frac{1}{\sqrt{5}}-\sqrt{\frac{3}{20}}+\frac{1}{\sqrt{10}}-\frac{1}{\sqrt{20}})[\frac{1}{4m_{u}m_{u}}-\frac{1}{m_{u}m_{d}}])
\end{split}
\end{equation}
The set of various combinations of Fock states
($|gg\rangle,|u\overline{u}g\rangle,|d\overline{d}g\rangle$) sea is
same for all baryon octet and decuplet members but there probability
distribution is different due to mass inherited from flavor leading
to different values of unknown parameters ($a_{0}, a_{8}, a_{10},
b_{1}$) etc.

\section{Estimation of hadron masses}
The baryon masses are calculated in literature using various models
and taking the inputs of constituent quark masses. The constituent
quark masses are model based parameters, so we allow suitable range
(in MeV) to them and tries to fit these parameters to the available
octet and decuplet masses, using {\it{mathematica 7.0}}. Here, the
input in the mass formulae are the coefficients calculated
statistically from the statistical model assuming sea quark-gluon
Fock states to be in specific flavor, spin and color states. The
model parameters for proton (in MeV) can be defined
as:\begin{equation*} m_{u}=m_{0}=290, m_{d}=340, b=600, a_{b}=220
\end{equation*}
Now substituting these values of parameters in equation (5.1), we
determine the mass of proton in D model with all sea contributions,
\begin{equation*} Mass_{(proton)} =937.6 MeV
\end{equation*} The mass of proton can be calculated with
other modifications such as in C model or with individual sea
contributions. Similarly, the model parameters for other baryons in
octet and decuplet are calculated and is shown in the form of
specific range below. For example, the model parameters (in MeV) for
$\Sigma^{*0}$ in decuplet can be shown as: \FloatBarrier
\begin{equation*}
m_{u}=m_{0}=260, m_{d}=310, m_{s}=450, b=600, a_{b}=200
\end{equation*} The calculation procedure described above leads to the
results for masses of other octet and decuplet baryons as presented
in table 3 and 4. The set of parameters (in MeV), for octet, in our
model are: b= 600 to 630 , $m_{u}$= 250 to 300 , $m_{d}$= 300 to
340, $m_{s}$= 450 to 550 and $a_{b}$= 200 to 230. The set of
parameters (in MeV), for decuplet, in our model are: b = 600 to 640,
$m_{u}$= 200 to 260, $m_{d}$= 250 to 330, $m_{s}$= 400 to 500 and
$a_{b}$ = 200 to 240. The masses of spin 1/2 and 3/2 particles are
computed in both C and D model with different sea contributions and
are shown in the table 3 and 4, respectively. The masses mentioned
in above set are the effective masses of quarks bound within hadrons
i.e constituent quark masses.
\setlength{\tabcolsep}{0.6em} %
{\renewcommand{\arraystretch}{1.6}%
\begin{table}[H]
\caption{Masses of octet particles}
\begin{center}
\begin{tabular}{|c|c|c|c|c|c|c|}
\hline
& \multicolumn{2}{p{2.2cm}|}%
{\centering C Model (MeV)}&\multicolumn{2}{p{2.2cm}|}%
{\centering D Model (MeV) \cite{13}}&\\
\cline{2-5}

\multicolumn{1}{|c|}{Particle $J^{P}=\frac{1}{2}^{+}$} &
\multicolumn{1}{p{2.1cm}|}{\centering With scalar sea}
&\multicolumn{1}{p{2.4cm}|}{\centering  With (scalar+tensor) sea}

&\multicolumn{1}{p{2.4cm}|}{\centering With (scalar+tensor) sea}
&\multicolumn{1}{p{2.3cm}|}{\centering  With
(scalar+vector\\+tensor) sea}


&\multicolumn{1}{c|}{Data (MeV)  \cite{13}}
\\
\hline \hline p&1044.48&1053.17&857.29&937.6&938.27\\\hline

n&1036.47&1045.7&938.8&939.85&939.56
\\\hline

$\Lambda$ &1175.6& 1187.53&1061.27&1113.5&1115.683\\\hline

$\Sigma^{+}$ &1249.53&1261.37&1141.98&1183.67&1189.37\\\hline

$\Sigma^{0}$ &1261.67&1239.81&1115.83&1189.05&1192.642\\\hline

 $\Sigma^{-}$ &1226.15&1237.78&1165.86 &1196.6&1197.449\\\hline

$\Xi^{0}$&1469.13&1469.13&1315.99&1314.89 &1314.86\\\hline

$\Xi^{-}$&1390.72&1403.2&1267.03&1321.21 &1321.71
\\\hline
\end{tabular}
\end{center}
\end{table}

\setlength{\tabcolsep}{0.6em} %
{\renewcommand{\arraystretch}{1.6}%
\begin{table}[H]
\caption{Masses of decuplet particles}
\begin{center}
\begin{tabular}{|c|c|c|c|c|c|c|}
\hline
& \multicolumn{2}{p{2.2cm}|}%
{\centering D Model (MeV)}&\multicolumn{2}{p{2.2cm}|}%
{\centering C Model (MeV)}&\\
\cline{2-5} \multicolumn{1}{|c|}{Particle $J^{P}=\frac{3}{2}^{+}$} &
\multicolumn{1}{p{2.1cm}|}{\centering With scalar sea}
&\multicolumn{1}{p{2.4cm}|}{\centering  With (scalar+tensor) sea}

&\multicolumn{1}{p{2.4cm}|}{\centering With (scalar+tensor) sea}
&\multicolumn{1}{p{2.3cm}|}{\centering  With
(scalar+vector\\+tensor) sea}

 &\multicolumn{1}{c|}{Data (MeV)  \cite{13}}
\\
\hline \hline

$\Delta^{++}$&1234.17&1207.44&1231.57&1230.62&1232.0\\\hline

$\Delta^{+}$&1234.87&1210.5&1231.87&1230.7&1232.0
\\\hline

$\Delta^{0}$ &1229.44&1206.91&1226.58&1231.45&1232.0\\\hline

 $\Delta^{-}$&1236.67&1204.38&1233.52& 1232.53&1232.0\\\hline

 $\Sigma^{*+}$
&1385.67&1364.76&1383.04&1382.17&1382.8\\\hline

$\Sigma^{*0}$ &1385.56&1367.32&1383.88&1383.22 &1383.7\\\hline

$\Sigma^{*-}$ &1391.33&1369.76&1388.62&1387.72&1387.2\\\hline

$\Xi^{*0}$&1621.9&1523.58&1531.46 &1531.09&1531.80\\\hline

$\Xi^{*-}$&1539.2&1516.11&1536.7&1535.8&1535.0
\\\hline

$\Omega^{-}$&1670.0&1650.7&1668.37&1668.04&1672.45
\\\hline
\end{tabular}
\end{center}
\end{table}
To check the validity of our results, we have checked Gellmann-okubo
mass formula for octet, equal spacing rule for decuplet and
electromagnetic mass splittings, with the results from our model.
The values obtained from our model are shown below every formula.
\begin{enumerate}
\item $\frac{N+\Xi}{2}=\frac{3\Lambda+\Sigma}{4}$\\
1129.40    1132.39

\item $\Delta-\Sigma^{*}=\Sigma^{*}-\Xi^{*}=\Xi^{*}-\Omega$\\
150.69  147.87  136.95
\item
$\Delta^{+}-\Delta^{++}$=
$n-p-(\Sigma^{+}+\Sigma^{-}-2\Sigma^{0})$\\
0.08  0.08

\item
$\Delta^{0}-\Delta^{+}$= $\Sigma^{*0}-\Sigma^{*+}$= $n-p$\\
0.75  1.05  2.25

\item
$\Delta^{-}-\Delta^{0}$= $\Sigma^{*-}-\Sigma^{*0}$=
$\Xi^{*-}-\Xi^{*0}$= $n-p+(\Sigma^{+}+\Sigma^{-}-2\Sigma^{0})$\\
1.08  4.5  4.34  4.42
\end{enumerate}

\section{Discussion and Conclusion}
We have calculated the masses of octet and decuplet particles using
the mass formulae consisting of constituent quark masses and
spin-spin interaction terms in statistical approach which assumes
sea to be an admixture of gluons and quark-antiquark pairs in
addition to valence quarks. The detailed analysis is based on
calculation of masses within different approaches namely C and D
model and further analyzing them by including the individual
contributions from scalar, vector and tensor sea. Here the sea with
spin 0, 1, 2 is called scalar, vector and tensor sea, respectively.
Model D find the probabilities of Fock states by suppressing the
contribution of states with higher multiplicities. Model D is
assumed to be a special case of Model C. To appreciate the
importance and validity of the sea with spin, various sea
contributions are presented in table 3 and 4. This individual
analysis of various sea and their possible contributions shows the
importance of scalar, vector and tensor sea in finding the masses of
spin $\frac{1}{2}$ and spin $\frac{3}{2}$ particles. Here, to check
the contribution from the pure scalar sea, the following assumptions
were made: $a_{0}\neq 0$ and $b_{1},b_{8},d_{1},d_{8} = 0$, for the
vector: $b_{1},b_{8}\neq0$ and $a_{0},d_{1},d_{8} = 0$ and similarly
for the tensor $d_{1},d_{8}\neq 0$ and $a_{0},b_{1},b_{8} = 0$. For
the case of the scalar plus tensor sea: $a_{0},d_{1},d_{8}\neq 0$
and $b_{1},b_{8} = 0$, for decuplet and similarly for octet.

Here, we can see from table 3 and 4 the masses comes out to be very
close to the data from D model in case of octet whereas from C model
in case of decuplet, respectively. Also specific dominancy of scalar
plus tensor sea contribution over vector is seen in both the cases.
Here, the coefficients associated with the scalar, vector and tensor
sea contributions are linearly related to the masses of octet and
decuplet with an effect of the interaction terms coming from
($a_{b}, b...$).

Here, the total multiplicities have been obtained using their
respective spin, color and flavor space. In case of octet, since D
model is giving better match with experimental data of masses and
that too, major contribution comes from scalar plus tensor sea that
dominates the contribution from vector sea by 66\%. Similarly, for
decuplet, C model gives better match with experimental data and
hence scalar plus tensor sea contribution dominates the vector sea
by 99.5\%. Hence, the calculation of masses for both
$J^{P}=\frac{1}{2}^{+}, \frac{3}{2}^{+}$, scalar plus tensor sea
contribution dominancy can be easily seen. In general, the sea is
found to be dynamic for the scalar and tensor in both octet and
decuplet. Here, spin-spin interaction term is dominating for
vectorial sea and hence it suppresses the overall contribution to
the masses from vector sea.

It can be very well seen from table 3, that the results from
contribution of scalar sea in C model is showing a deviation of 5\%
to 11\% while combination of scalar-tensor contribution shows a
deviation of 3\% to 12\% when compared with experimental values
available for $\frac{1}{2}^{+}$ particles . On the other hand
results from D model are deviating 0.03\% to 0.47\%, when total sea
is contributing, which shows that masses from total sea is providing
good match with PDG data. Similarly, it can be seen from table 4,
for decuplet, that contribution from total sea in case of C model is
providing a good match with experimental values available as
compared to individual contribution from sea.

It is also interesting to note  that the set of constituent quark
masses are different for every particle in octet and decuplet. If we
insist to take single set of quark masses to hold for all particles
and vary these masses, the overall best fit to the hadron data is
deviating more from the experimental data. Moreover, the particles
with two or three heavy (strange) quarks leads to less or negligible
contribution of spin-spin interaction term to the overall mass of
particles and hence will be less significant.

\section*{Acknowledgement}
The authors gratefully acknowledge the financial support by the
Department of Science and Technology (SB/FTP/PS-037/2014), New
Delhi.

\end{document}